\documentclass[prd,twocolumn,amsmath,amssymb,floatfix,nofootinbib,superscriptaddress]{revtex4}
\pdfoutput=1

\usepackage{graphicx}
\usepackage{bm}
\usepackage{amsmath,amssymb}
\usepackage{epsfig}
\usepackage{color}
\usepackage{natbib}
\usepackage{enumerate}

\usepackage[pdftex]{hyperref}
\usepackage{ifthen}
\usepackage{subfigure}
\usepackage{hhline}
\usepackage{multirow}

\newcommand{\planck}{\textit{Planck}}

\newcommand{\rd}{r_\text{d}}
\newcommand{\rdhat}{\hat{r}_\text{d}}

\newcommand{\unit}[1]{\;\text{#1}}
\newcommand{\fig}[1]{Fig.~\ref{#1}}
\newcommand{\zdrag}{z_\text{drag}}
\newcommand{\hrd}{\hat{r}_\text{d}}

\newcommand{\vxi}{\bm{\xi}}
\newcommand{\vxihat}{\hat{\bm{\xi}}}
\newcommand{\chimps}{h^{-1} \text{Mpc}}

\newcommand{\Mpc}{\text{Mpc}}

\newcommand{\Hunit}{~\text{km}~\text{s}^{-1} \Mpc^{-1}}

\providecommand{\CAMB}{\textsc{camb}}
\providecommand{\COSMOMC}{\textsc{CosmoMC}}

\providecommand{\LCDM}{{$\rm{\Lambda CDM}$}}

\newcommand{\begm}{\begin{pmatrix}}
\newcommand{\enm}{\end{pmatrix}}

\newcommand\ba{\begin{eqnarray}}
\newcommand\ea{\end{eqnarray}}
\newcommand\bea{\begin{eqnarray}}
\newcommand\eea{\end{eqnarray}}

\newcommand\be{\begin{equation}}
\newcommand\ee{\end{equation}}

\newcommand{\ud}{{\rm d}}

\newcommand{\mC}{\bm{C}}

\def\eprinttmp@#1arXiv:#2 [#3]#4@{
\ifthenelse{\equal{#3}{x}}{\href{http://arxiv.org/abs/#1}{#1}}{\href{http://arxiv.org/abs/#2}{arXiv:#2} [#3]}}

\providecommand{\eprint}[1]{\eprinttmp@#1arXiv: [x]@}
\newcommand{\adsurl}[1]{\href{#1}{ADS}}
\providecommand{\bibinfo}[2]{\ifthenelse{\equal{#1}{isbn}}{
\href{http://cosmologist.info/ISBN/#2}{#2}}{#2}}

\begin{document}

\title{Accuracy of cosmological parameters using the baryon acoustic scale}

\author{Kiattisak Thepsuriya}
\affiliation{Department of Physics \& Astronomy, University of Sussex, Brighton BN1 9QH, UK}

\author{Antony Lewis}
\homepage{http://cosmologist.info}
\affiliation{Department of Physics \& Astronomy, University of Sussex, Brighton BN1 9QH, UK}


\begin{abstract}
Percent-level measurements of the comoving baryon acoustic scale standard ruler can be used to break degeneracies in parameter constraints from the CMB alone. The sound horizon at the epoch of baryon drag is often used as a proxy for the scale of the peak in the matter density correlation function, and can conveniently be calculated quickly for different cosmological models. However, the measurements are not directly constraining this scale, but rather a measurement of the full correlation function, which depends on the detailed evolution through decoupling. We assess the level of reliability of parameter constraints based on a simple approximation of the acoustic scale compared to a more direct determination from the full numerical two-point correlation function. Using a five-parameter fitting technique similar to recent BAO data analyses, we find that for standard $\Lambda$CDM models and extensions with massive neutrinos and additional relativistic degrees of freedom, the approximation is at better than $0.15\%$ for most parameter combinations varying over reasonable ranges.
\end{abstract}
\maketitle

\section{Introduction}
\label{intro}

Observations of the microwave background and large-scale structure can be used to measure a variety of cosmological parameters. With accurate measurements of the distance-redshift relation, we can also start to probe the nature of dark energy. One of the cleanest probes of this is the baryon acoustic oscillation (BAO) signal imprinted in the power spectrum of the large-scale clustering of galaxies (see Ref.~\cite{Bassett:2009mm} for a review).

The scale of acoustic oscillations in the CMB has now been measured to 0.1\%~\cite{Ade:2013zuv}, and large-scale galaxy surveys have now been able to measure BAO and hence the baryon acoustic scale at lower redshift to approximately 1\%~\cite{Anderson:2013zyy}. A combination of these data gives a powerful probe of cosmological parameters, constraining both early and late-universe physics, geometry, and evolution. Parameters which are subject to the geometric degeneracy (such as dark energy, curvature, and light massive neutrinos) when measured from the single source-plane CMB can be constrained much better using BAO, and hence our current best knowledge of these and other parameters critically depends on a reliable physical interpretation of the baryon acoustic scale.

The aim of this paper is to test to what level the standard proxy for the acoustic scale measured in the correlation function of galaxies is accurate, justifying the use of this approximation in extended cosmological models, for example in the analysis of BAO measurements combined with Planck~\cite{Ade:2013zuv}. Our analysis is especially timely because the renewed interest in extended neutrino models as a possible way of partially reconciling some of the apparent conflicts between existing data in the $\Lambda$CDM model, and using cosmology to tightly constrain possible solutions to neutrino reactor anomalies~\cite{Wyman:2013lza,Leistedt:2014sia,Battye:2013xqa,Hamann:2013iba,Vincent:2014rja,MacCrann:2014wfa,Battye:2014qga}. Models with additional relativistic degrees of freedom typically significantly modify the expansion history and evolution before recombination, and hence potentially significantly modifying the shape of the matter correlation function compared to the standard assumption of $\Lambda$CDM. If the simple acoustic scale approximation starts to break down, a more complete direct analysis of the correlation function data would likely be required.

In Sec.~\ref{baoscale} we start by a review of the baryon acoustic scale, how it can be defined and related to measurements of galaxy clustering. Sec.~\ref{method} then describes how we test the acoustic scale approximation by comparing constraints obtained directly from the correlation function to those obtained using the approximation. We only aim to test the accuracy of theoretical modelling, and hence consider simplified and idealized data. This should be sufficient to test whether the single acoustic scale number is sufficient to accurately capture the parameter dependence of the shape of the correlation function when fit with a scaling parameter (as in recent BAO measurements). Finally in Sec.~\ref{results} we give results for various parameter variations of standard and extended cosmological models, and end with our conclusions.

\section{The baryon acoustic scale}
\label{baoscale}

\begin{figure*}
\centering
\includegraphics[width=7in]{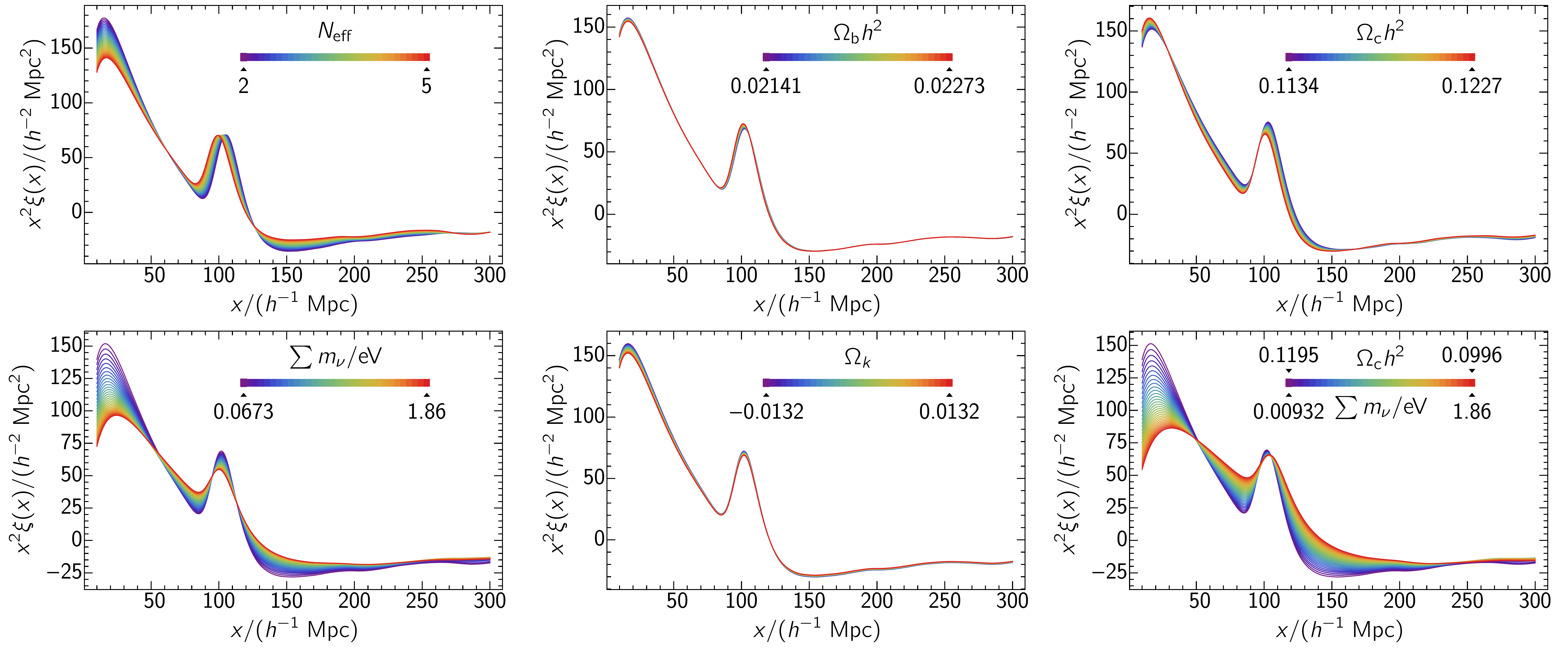}
\caption{Theoretical predictions for the two-point matter density correlation function $\xi(x)$ at redshift $z = 0$. Colours show how the function changes as different cosmological parameters are varied away from the fiducial flat \LCDM\ model. In all plots, $H_0$ is fixed at its fiducial value, and only one cosmological parameter is varied at a time, except in the last plot where $\Omega_c$ and $\Omega_\nu$ are simultaneously varied to keep $(\Omega_c + \Omega_\nu) h^2$ fixed at $0.1196$. Apart from the fourth (in which only $\sum m_\nu$ is varied) and last plots, the total neutrino mass is fixed at zero. Note that in general both the shape and peak position change.}
\label{figcor}
\end{figure*}

In the pre-recombination plasma, Thomson scattering kept the photons and baryons coupled together. Acoustic oscillations in the photon-baryon fluid imprint the acoustic oscillations observed in the CMB power spectra, as well as a smaller acoustic oscillation in the distribution of the total matter density after recombination due to the oscillations imprinted in the baryon component. The distance that the photon-baryon acoustic wave can travel (sound horizon) at the time of photon decoupling defines the scale of acoustic oscillations seen in the CMB power spectra. Since there are far more photons than baryons, after photon decoupling the photons continued to drag baryons with them slightly longer into the Compton drag epoch. The sound horizon when the baryons stop being dragged around by the photons therefore roughly defines the baryon acoustic scale, which is slightly larger than the sound horizon seen in the CMB.

From a theoretical model, the baryon acoustic scale is usually defined as the comoving sound horizon at the baryon drag epoch, $\rd \equiv r_\text{s} (\zdrag)$, where $\zdrag$ is the redshift at which the baryon velocity decouples from the photons. The sound horizon can be calculated by integrating the speed of the photon-baryon fluid,
\begin{equation}
  r_{\rm s}(z) = \int_0^{\eta(z)}  \frac{d\eta^\prime}{\sqrt{3(1+R)}},
\end{equation}
where $R$ is given in terms of the baryon density $\rho_{\rm b}$ and photon density $\rho_\gamma$ by $R \equiv 3 \rho_{\rm b}/(4\rho_\gamma)$ and $\eta$ is the conformal time. After $\zdrag$ the baryon perturbations stop undergoing acoustic oscillations and start to grow with the dark matter perturbations, but the process of decoupling is in reality gradual, so $\zdrag$ is defined to be an indicative central value where
$\tau_{\rm d} \sim 1$, where~\citep{Hu:1995en}
\begin{equation}
  \tau_{\rm d}(\eta) \equiv \int^\eta_{\eta_0} \ d\eta^\prime (\partial_\eta \tau)/R.
\end{equation}
Here $\tau$ is the Thomson optical depth from recombination (without reionization) and $\eta_0$ the conformal time today. For a given set of cosmological parameters, and model for the recombination history, $\rd$ can be calculated quickly numerically in terms of background quantities, and is a standard output of \CAMB~\cite{Lewis:1999bs}. For standard cosmological parameters $\rd$ can be predicted quite accurately using approximate fitting functions~\cite{Eisenstein:1997ik}, however for precision measurements, and generality in extended models, we use a direct numerical calculation as in most recent analyses.

Observationally, the baryon acoustic scale is the comoving scale corresponding to the position of the peak of the density correlation function, which can (indirectly) be measured from observations of galaxy clustering. The full density correlation function can be predicted accurately using linear perturbation theory. However due to issues of bias, non-linearities and observational complications, it cannot easily be measured directly in galaxy surveys. However the position of the peak of the correlation function is thought to be much more robust, and therefore a powerful probe of the underlying cosmological model~\cite{Eisenstein:1998tu,Eisenstein:2005su}. In practice, comoving distances cannot be observed directly (only angular scales and redshifts), so the observed angular correlation function at redshift $z$ is related to the underlying comoving acoustic scale by the comoving angular diameter distance $(1+z)D_{\rm A}(z)$ (for perturbations orthogonal to the line of sight), and the Hubble parameter $H(z)$ (for perturbations along the line of sight). A spherically averaged distance $D_\text{V} (z)$ is often defined by
\begin{equation}
\label{dv}
D_\text{V} (z) \equiv [cz (1 + z)^2 D_\text{A}^2 (z) H^{-1} (z)]^{1/3},
\end{equation}
though it is now becoming possible to place constraints on the radial and transverse parts separately~\cite{Anderson:2013zyy}.

The background functions $D_{\rm A}(z)$,  $H(z)$ and $D_{\rm V}(z)$ can be calculated easily from a given cosmological model, so the observed scale of the peak in the correlation function measures the parameter combination $\rd/D_\text{V}(z)$ (and equivalently for the radial and transfer components if they are resolved separately; we use this particular combination as a concrete example below). Often $\rd$ is considered to be accurately known by fits to CMB observations, in which case most of the information is about the late-time geometry. However in models with extra relativistic degrees of freedom (for example, sterile neutrinos), massive neutrinos, or other extensions of the \LCDM\ model, $\rd$ can also vary, and it is important to account for the model dependence of both parts of the ratio~\cite{Hamann:2011ge}.

An optimal analysis of galaxy clustering data would model the full shape dependence of the correlation function, marginalizing appropriately over uncertainties in the bias, non-linearities and observational systematics, and determine parameter constraints directly from the theoretically predicted density correlation function in different models. A `BAO-only' measurement attempts to abstract from such an analysis a measured peak correlation scale which can then be used as a kind of radical data compression when comparing with theoretical models, which is simple to interpret independently of the underlying modelling assumptions.
There is however considerable freedom in how precisely the scale of the peak of the correlation function is defined from observed galaxy data. To be specific we follow Refs.~\cite{Xu:2012hg,Anderson:2013zyy,Ross:2014qpa}, defining a scale parameter $\alpha$ by fitting the galaxy data (power spectrum or two-point correlation)
relative to some scaled fiducial cosmological prediction with the assumption that
\begin{equation}
\label{assumption}
D_\text{V} / \rd \propto \alpha.
\end{equation}
The proportionality constant is to be determined from the fiducial model.

The fitting of the redshift-space two-point correlations $\xi(x_i)$'s is done by minimizing
\begin{equation}
\label{chi2}
\chi^2 = (\vxi_\text{fit} - \vxihat)^\mathrm{T} \mC^{-1} (\vxi_\text{fit} - \vxihat),
\end{equation}
where $\vxihat$ is a vector of the observed galaxy correlation function estimates for bins centred at $x_i$, $\vxi_\text{fit}$ consists of the values of a fitting function at the bin centres, and $\mC$ is the covariance matrix that specifies the error model associated with $\vxihat$. Recent BAO data analyses from the clustering of galaxies~\cite{Anderson:2013zyy} put five parameters in the fitting function, chosen to be in the form of
\begin{equation}
\label{xifit}
\xi_\text{fit} (x) = B^2 \xi_\text{f} (\alpha x) + A(x)
\end{equation}
where $\xi_\text{f} (x)$ is an appropriate fiducial template correlation function, $B^2$ is a constant that accounts for an overall bias, and
\begin{equation}
A(x) = \frac{a_1}{x^2} + \frac{a_2}{x} + a_3.
\end{equation}
The smoothly-varying $A(x)$ function (with three free parameters, $a_1$, $a_2$ and $a_3$) accounts for the unknown overall shape of the galaxy correlation, and is intended to remove most of the dependence on the unknown scale-dependence of the bias and non-linear physics.
The assumption \eqref{assumption} relates the acoustic scale to the quantity actually gained from measurements, the value of the scale dilation parameter $\alpha$ that minimizes the $\chi^2$. Extensive tests have investigated the reliability of this kind of procedure from the data side in simple models (e.g. Ref.~\cite{Magana:2013wpa}). In this paper, we assess the accuracy of the assumed theoretical $\rd$ dependence of Eq.~\eqref{assumption} across different extended cosmological models.

Examples of how the two-point density correlation functions change with cosmological parameters are shown in \fig{figcor}. Previous cosmological analyses, including the main \planck\ cosmological parameter analysis~\cite{Ade:2013zuv}, have assumed that the parameter dependence of $\rd$ accurately models the change in peak position as reported from the data via the $\alpha$ scale parameter as defined above. Although it is clearly qualitatively correct, with precision now reaching the sub-percent level, it is not immediately obvious to what level of accuracy this assumption is valid, and at what point a more detailed modelling of the change of correlation function shape will be required; answering these questions is the purpose of this paper.

\section{Methodology}
\label{method}

\begin{figure}
\centering
\includegraphics[width=3.4in]{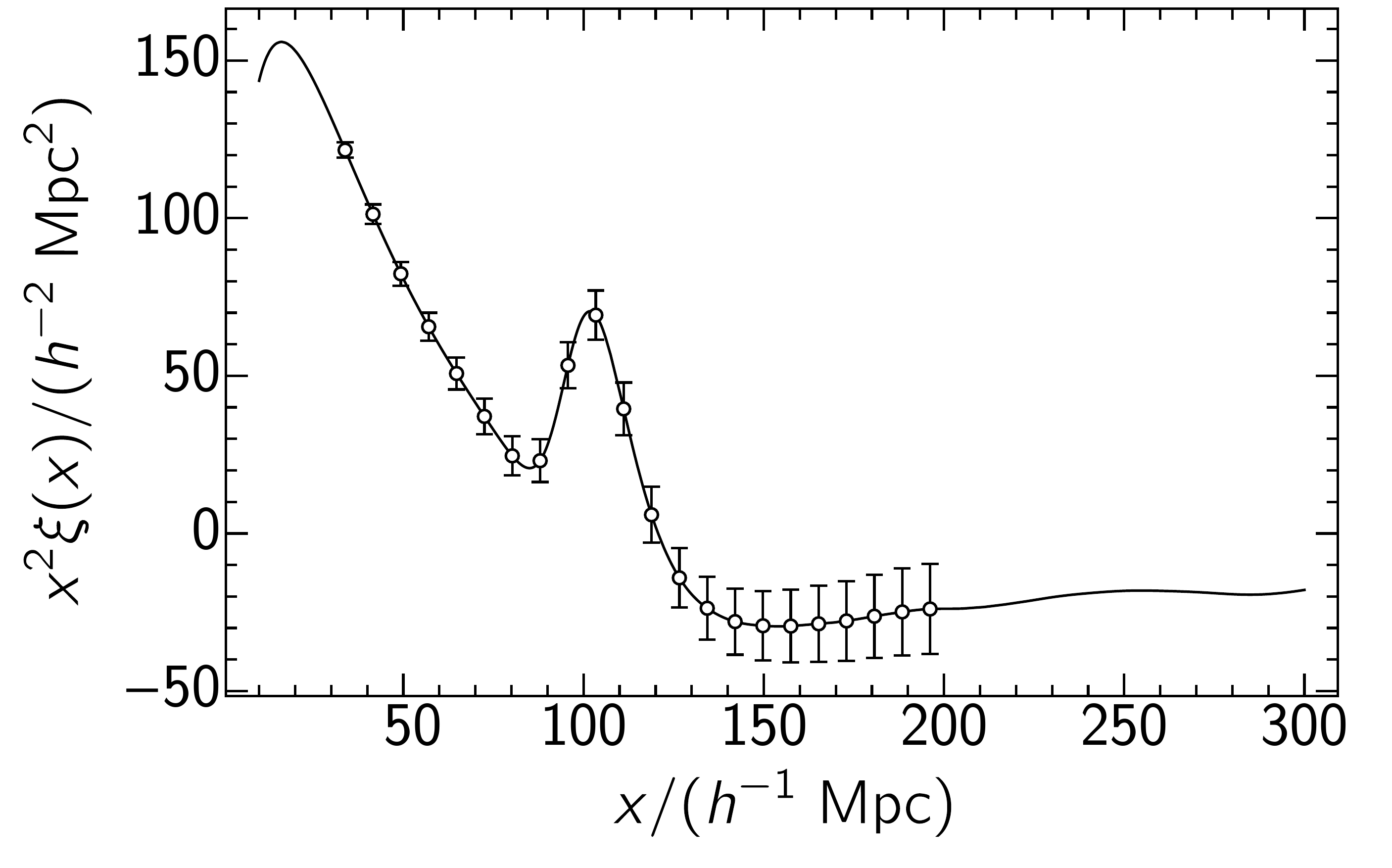}
\caption{The two-point correlation of the fiducial model calculated for $(10 < x < 300) \chimps$ from the matter power spectrum. Error bars show the diagonal components of the error model about this fiducial function sampled for 22 bins of width $7.73 \chimps$, starting at $x_1 = 33.86 \chimps$. The fake `data' in each model is simulated for the bin centres; when fitting a scaled fiducial fitting function the correlation function is interpolated to the bin positions.}
\label{figfid}
\end{figure}

We now outline our procedure for testing whether using the observed acoustic scale parameter $\alpha$ with a numerical calculation of $\rd$ for each cosmological model accurately recovers the parameter dependence of the underlying data. Rather than being specific to a particular data set, we use fake `data' generated with different underlying cosmological parameters, with a fixed error model that approximates the error model of recent BAO results.

Measurements of the acoustic scale can be made at different redshifts, and the effective distance $D_\text{V}(\tilde{z})$ for each redshift bin around an effective redshift $\tilde{z}$ can easily be computed from Eq.~\eqref{dv}. Here we are not concerned by the late-time geometry dependence of the result, and instead focus on whether the parameters that affect the acoustic scale at recombination are being adequately modelled. We therefore focus on a single fiducial redshift, and expect our conclusions to be very independent of that assumption because the shape of the correlation function in comoving distance is nearly constant long after recombination when it is observed.

Specifically we test to what accuracy the assumption that $\rd \propto 1/\alpha$ holds by evaluating a simple estimate
\begin{equation}
\label{prop}
\hrd  = r_\text{fid} \alpha_\text{fid} / \alpha.
\end{equation}
For various different theoretical models we make a fake data vector $\vxihat$, and then fit a fiducial-model correlation function $\xi_\text{f} (x)$ scaled by $\alpha$ as in Eq.~\eqref{xifit}. The best-fit value of alpha in each model gives the estimate $\hrd$ of the value of the acoustic scale $\rd$ in that model, which we then compare to the true value. By `simple' we mean that inputs for Eqs.~\eqref{chi2} and \eqref{xifit} are provided as follows:
\begin{enumerate}
\item The components of $\vxihat$ for each cosmological model are given by a numerical evaluation of the two-point correlation function
\begin{equation}
\label{correlator}
\hat{\xi}(x_i) = \int \mathrm{d}k \frac{k^2}{2\pi^2} P_\text{g} (k) j_0 (kx_i),
\end{equation}
where $P_\text{g} (k)$ is a simplistic model of the theoretical galaxy power spectrum described further below, $j_0 (kx)$ is the zeroth-order spherical Bessel function, and $x_i$ is the bin centre. To avoid having to do many simulations we fix the data points to their expected values, with no scatter.

\item The template function is chosen to be the full numerical $\xi_\text{fid} (x)$ constructed from  $P_\text{g,f} (k)$ of the fiducial model. The quantity $\alpha_\text{fid}$ in Eq.~\eqref{prop} consequently becomes unity as it results from fitting $\xi_\text{fid} (x)$ against a binned version of itself.
\item The covariance matrix is fixed for all cosmological models, and is estimated using the binned Gaussian covariance matrix derived in Ref.~\cite{Xu:2012hg}:
\begin{align}
\label{covmat}
&C_{ij} = \frac{2}{V} \int \frac{k^2\ud k}{2\pi^2} \Delta j_1 (kx_i) \Delta j_1 (kx_j) [P_\mathrm{c} (k) + \mathcal{N}]^2 \nonumber\\
\quad&
\end{align}
where
\begin{equation}
\Delta j_1 (kx_i) = 3 \frac{x_{i,2}^2 j_1 (kx_{i,2}) - x_{i,1}^2 j_1 (kx_{i,1})}{k(x_{i,2}^3 - x_{i,1}^3)},
\end{equation}
$j_1 (x)$ is the first-order spherical Bessel function, $x_{i,1}$ is the left edge of the bin at centre $x_i$, $x_{i,2}$ is the right edge, $V$ is the comoving volume of the survey from which the matter density data is obtained, $P_\mathrm{c} (k)$ is an appropriate form of the power spectrum that captures observational variance information (redshift-space distortions, etc), and $\mathcal{N}$ is the shot-noise error from finite galaxy density.
\item $P_\text{c} (k)$ is chosen simply to be the fiducial-model power spectrum $P_{g,f}(k)$, and $\mathcal{N}$ the linear Poisson shot-noise without $z$ dependence, which is equal to the inverse of the number density of galaxies in the survey~\cite{Meiksin:1998mu}. This means that deviation from an ideal observation is only minimally represented, and hence the test is free from the particulars of how the noise, redshift-space distortions, and so on are modelled.
\end{enumerate}

For each fake data vector $\vxihat$, the bin centres $x_i$ are chosen similarly to Ref.~\cite{Anderson:2013zyy}, with the range $(30 < x < 200) \chimps$ divided into $22$ bins of width $7.73 \chimps$. For the fit $\vxi_{\rm fit}$, we extend the distance range of the fiducial correlation function to $(10 < x < 300)\chimps$ to accommodate the variation of $\alpha$ during minimization (confined to the range $0.6 < \alpha < 1.4$). The scaled fiducial correlation function is evaluated at the position of the bin centres by interpolation; see~\fig{figfid} for what some typical fake `data' looks like.

For all correlation functions and $C_{ij}$, we use a simple model for the galaxy power spectrum $P_\text{g}(k)$ in the form
\begin{equation}
P_\text{g}(k) = b^2 P(k) e^{-k^2 a^2},
\end{equation}
where $P(k)$ is the theoretical matter power spectrum from \CAMB. The multiplicative constant $b^2$ accounts for the large-scale bias, and the Gaussian damping term helps numerical convergence of the Bessel transform to the correlation function. The bias $b^2$ is chosen to be $4$, which does not affect the acoustic scale~\cite{Mehta:2011xf} since its effect on the fitting can be absorbed by the fitting parameter $B$. Choosing $a = 1 \chimps$ is sufficiently below the scale of interest that the acoustic peak of $\xi(x)$ is not significantly affected by the damping.

For the fiducial cosmology we choose $\Omega_\text{b} h^2 = 0.02207$, $\Omega_\text{c} h^2 = 0.1196$, $\Omega_K = 0$, $\sum m_\nu = 0 \unit{eV}$, $H_0 = 67.4 \Hunit$, and $\Omega_m = 0.311859$, for which \CAMB\ gives $r_\text{d,fid} = 147.56 \unit{Mpc}$. To calculate the comoving volume in the covariance matrix, we also need to specify the redshift range of the galaxy survey data, and the sky coverage of the survey (if not full-sky). Together with the background number density of galaxies, we have three survey-specific numbers entering the test via our assumed covariance matrix. We choose them to be those of a recent baryon oscillation spectroscopic survey of galaxies~\cite{Anderson:2013zyy}, with $0.15 < z < 0.7$, $8377 \unit{deg}^2$ coverage, and $1009172$ galaxies. With these numbers, the comoving volume given our fiducial cosmology is $4.45 h^{-3} \unit{Gpc}^3$ and consequently $\mathcal{N}$ is $4407 h^{-3} \unit{Mpc}^3$.

By having $\xi_\text{fid} (x)$ as the template function parameterized as in Eq.~\eqref{xifit}, it may happen that $\chi^2$ is minimized with the fitting function being essentially $A(x)$, with $B^2$ taking a very small value. To prevent this, a Gaussian prior is put on $\log(B^2)$ with a mean of $0$ and standard deviation of $0.4$, following Ref.~\cite{Xu:2012hg}. This has a negligible effect on the minimization for models reasonably close to the fiducial model.

We compare $\hrd$ from Eq.~\eqref{prop} against the accurate numerical value of $\rd$ calculated by \CAMB\ for models that are mostly one-parameter deviations from the fiducial model. Various parameter variations are considered:
\begin{enumerate}
\item $2 < N_\text{eff} < 5$
\item $0.02141 < \Omega_\text{b} h^2 < 0.02273$
\item $0.1134 < \Omega_\text{c} h^2 < 0.1227$
\item $-0.0132 < \Omega_K < 0.0132$
\item $0.00932\unit{eV} < \sum m_\nu < 1.86\unit{eV}$ (assuming three degenerate masses with $N_\text{eff}$ fixed at $3.046$)
\item As (5) but with $(\Omega_\nu + \Omega_\text{c}) h^2$ kept fixed at $0.1196$ by varying $\Omega_\text{c} h^2$
\item Simultaneous variation of $N_\text{eff}>3.046$ and $m_\nu$ of a single type of massive sterile neutrinos
\end{enumerate}
The ranges chosen are motivated by $95\%$ limits from \planck\ (though here we do not vary other parameters along CMB degeneracy directions, so for fixed values of other parameters the range is significantly broader than allowed by \planck). The range of extended neutrino parameters is motivated by the range that may conceivably be relevant for combined fits with other data. The curvature parameter $\Omega_K$ is expected to have only a very small effect on $\rd$ since it only has a significant effect on the evolution of the universe at late times, and we can check that this is indeed the case.

\section{Results}
\label{results}

\begin{figure*}
\centering
\includegraphics[width=7in]{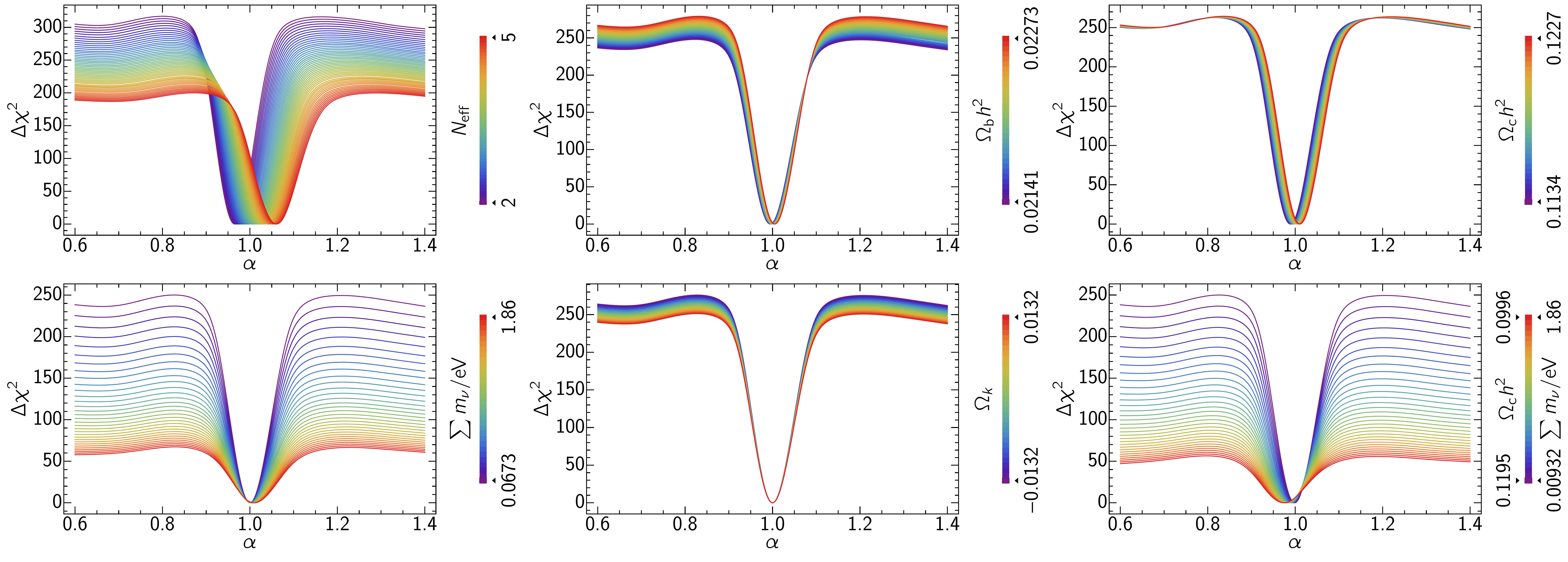}
\caption{Values of $\Delta \chi^2(\alpha)$ for fits using various values of underlying parameters (colours), each of which is computed by optimizing $B^2$, $a_1$, $a_2$, and $a_3$ for each value of $\alpha$ in the range of $1 \pm 0.4$. This is only to visualize the optimization of $\alpha$; in practice all five fitting parameters are optimized simultaneously.}
\label{figchi2}
\end{figure*}

\begin{figure*}
\centering
\includegraphics[width=7in]{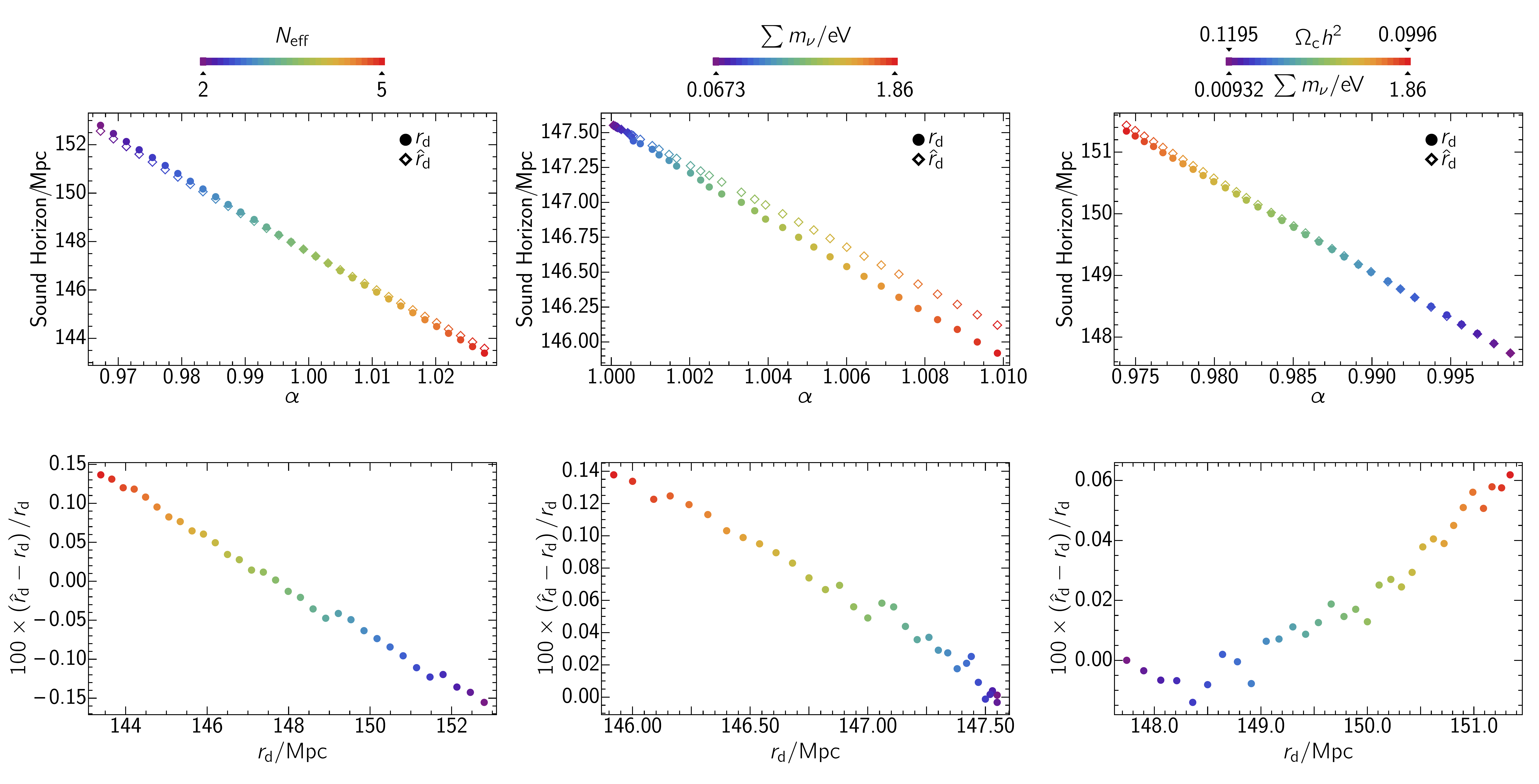}
\caption{The upper three plots show the acoustic scale parameter $\rd$ plotted against the scale parameter $\alpha$ at which $\chi^2$ is minimized for various values of cosmological parameters. The estimated $\hat{r}_\text{d}$ is derived from the corresponding best-fit $\alpha$, and compared to the exact sound horizon $\rd$. Generally close agreement between the lines shows that the approximation captures the main dependence well. The lower three plots show the percentage deviation of the estimated acoustic scale $\hat{r}_\text{d}$ from the exact $\rd$ plotted against $\rd$ for various extension to the \LCDM\ model. The plots show that the approximation of Eq.~\eqref{prop} is accurate to within $0.15\%$ over broad parameter ranges. Some numerical noise is visible, at a level that does not affect our conclusions.}
\label{figrd}
\end{figure*}

\fig{figchi2} shows the variation of $\chi^2$ used for the optimization of $\alpha$ as various parameters are varied, with $\Delta\chi^2 (\alpha) \equiv \chi^2(\alpha) - \chi^2_\text{min}$. Using the best-fit value of $\alpha$ for each model, \fig{figrd} shows the corresponding approximation for the acoustic scale $\rdhat$ calculated assuming $\rd \propto 1/\alpha$ from Eq.~\eqref{prop}, compared to exact $\rd$. This shows that as expected the approximation is correctly capturing the main change in the acoustic scale. In more detail, \fig{figrd} shows the percentage error compared to the exact result, which shows that overall the approximation is accurate to within about $0.15\%$ for our reasonably broad ranges of cosmological parameters. This is below the current accuracy of the measurement of the comoving BAO acoustic scale, which is at approximately the $1\%$ level~\cite{Anderson:2013zyy}, though not by a large factor.

In all cases, $A(x)$ is close to zero throughout the fitting range. Setting it to zero only negligibly affects the results. On the other hand, $B^2$ is not always close to one in some of the cases (varying $\sum m_\nu$, varying $\Omega_\text{c} h^2$, and varying both the temperature and mass of one type of massive sterile neutrinos). The amplitude of the BAO peak given by a particular model can be different to that of the template $\xi_\text{f} (x)$, and the $B^2$ amplitude parameter is then indispensable.

We briefly comment on the accuracy of the approximation in the various cases as follows.

\subsection{\LCDM\ parameters}
\label{LCDM}

As shown in \fig{figcor}, varying $\Omega_\text{b}$, $\Omega_\text{c}$, and $\Omega_K$ within the range of the {\planck} error bars does not move the models very far away from the fiducial one. The sound horizon remains accurately at the fiducial value regardless of change in $\Omega_K$, as expected. For $\Omega_\text{b} \pm 2\sigma$ and $\Omega_\text{c} \pm 2\sigma$, the approximation is very accurate (within $0.03\%$).

\subsection{Extra relativistic energy density}
\label{Neff}

The Standard Model has three flavours of active neutrinos, giving the effective number of massless neutrinos $N_\text{eff}=3.046$~\cite{Mangano:2005cc}. The possible presence of additional massless sterile neutrinos is usually quantified by the change in the effective number given by $\Delta N_\text{eff}$. The first plot of \fig{figrd} shows that the acoustic scale approximation is still robust to $0.15\%$ accuracy as $\Delta N_{\text{eff}}$ is varied, up to even $\Delta N_\text{eff} \approx 2$, which is well outside the range constrained by {\planck} and lensing observations~\cite{Battye:2013xqa,Dvorkin:2014lea}.

\subsection{Massive neutrinos}
\label{MassiveNU}

\begin{figure}
\centering
\includegraphics[width=3.4in]{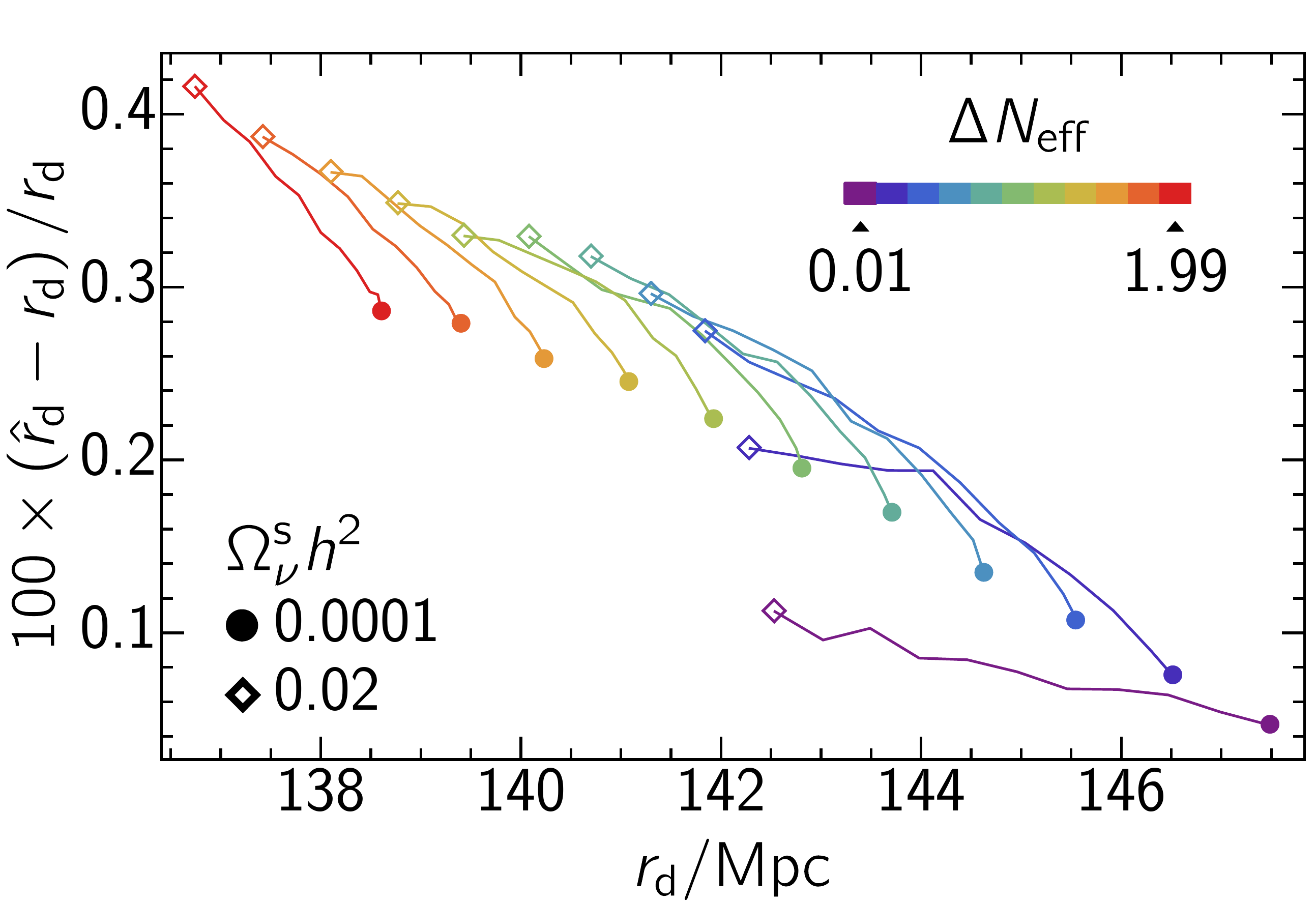}
\caption{The deviation percentage of the estimated sound horizon $\hat{r}_\text{d}$ from the exact sound horizon $\rd$ plotted against $\rd$, in case when both $N_\text{eff}$ and $\sum m_\nu$ are varied, given three types of active neutrinos and one type of sterile neutrinos. Two types of active neutrinos are set to be massless, and the mass of the massive one is fixed at $0.06\unit{eV}$. $N_\text{eff}$ is set to range from $3.046+0.01$ to $3.046+1.99$. Each line corresponds to a particular value of $N_\text{eff}$ and the variation of the mass of sterile neutrinos in the range $0.00941 < \sum m_\nu^\text{s} < 1.88\unit{eV}$. The plot shows that \eqref{prop} is accurate to within $\sim 0.4\%$ even in relatively extreme cases.}
\label{figrdnu}
\end{figure}

As neutrino data from various sources have now confirmed that neutrinos are in fact not massless~\cite{GonzalezGarcia:2012sz}, it is more important to look at the response of $\rd$ as $\sum m_\nu$ changes. With $N_\text{eff}$ fixed at $3.046$, the second plot in \fig{figrd} shows that the fitting approximation is accurate to within $0.15\%$ up to large masses.

At late time of cosmic evolution, cold dark matter behaves like massive neutrinos. We therefore also explore the accuracy of the assumption were the universe composed of less cold dark matter but more massive neutrinos. We fix $(\Omega_\text{c} + \Omega_\nu) h^2$ at the fiducial value of $\Omega_\text{c} h^2$, and find that the approximation is still accurate to within $0.1\%$, as shown by the third plot of \fig{figrd}, even with large values of $\sum m_\nu$.

Sterile neutrinos, if present, may also be massive, and this possibility has attracted significant attention recently as a possible way to reduce some of the tensions between different datasets. \fig{figrdnu} shows the accuracy of the acoustic scale approximation when both $\Delta N_{\rm eff}$ and $m_\nu$ are varied to allow for the possibility of one type of massive sterile neutrinos; this shows that the approximation is still accurate to within about $0.4\%$, within a range much broader than proposed values constrained by {\planck}, lensing and cluster observations~\cite{Battye:2013xqa,Dvorkin:2014lea}.

\section{Conclusions}

We have shown that the comoving sound horizon at the baryon-drag epoch, fit from data by a simple scaling from a fiducial model, can be accurately modelled using a numerical calculation of the acoustic scale $\rd$. Given reasonable ranges of values of cosmological parameters, the agreement is within $0.15\%$, which is sufficient for current observations. This validates results obtained in extended models when combining CMB, BAO and other data using the simple acoustic scale approximation as implemented for example in \COSMOMC~\cite{Lewis:2002ah} and used for example by Refs.~\cite{Ade:2013zuv,Battye:2013xqa,Wyman:2013lza,Leistedt:2014sia,Battye:2014qga}. Wider parameter variations give worse agreement, but even in a relatively extreme cases, such as the standard model with a high density of massive sterile neutrinos, the accuracy should still be well within a percent, demonstrating good tolerance of the assumption \eqref{assumption}. As emphasized by Ref.~\cite{Hamann:2011ge}, for results to be valid it is important to model the cosmological parameter dependence of $\rd$ (which has previously sometimes been neglected entirely), but a fast numerical calculation is sufficient for a good correspondence with a much more time consuming full correlation function fitting procedure

Future observations will continue to improve the accuracy of BAO measurements, as surveys measure larger areas to greater depth, and analysis methods improve. Although our results are based on comparison to fits with the accuracy of current data, the acoustic scale approximation is sufficiently accurate that it is likely to remain valid for most extended models in the immediate future, as BAO measurements carried out by, for example, eBOSS, DES, HETDEX, and WFIRST will still be in the percent-level regime~\cite{Percival:2013awa,2013MNRAS.431.1777G,Green:2012mj}.
The accuracy of Euclid is expected to be within 0.4\% in many bins between $0.7 \leq z \leq 1.7$~\cite{Sutherland:2012ys}, and LSST within 0.5\% in ten bins between $0.3 \leq z \leq 3$~\cite{Zhan:2008jh}, so ultimately accuracy of $\sim 0.1\%$ may be required when the full combined redshift range is considered. However by the time sub-percent-level measurements are available, cosmological parameters, especially those related to neutrinos, will probably have already been further constrained to a relatively small region of parameter space, and the assumption \eqref{assumption} is likely to remain valid within that range, but will require further validation.

As precision increases errors in the fitting approximation may become more important, and will need to be tested using the error model and fitting procedure used at the time. It is also be possible to use more information, rather than compressing the observed correlation down into just a few acoustic scale measurements. In principle, the fitting procedure for non-linearities, bias and systematics could easily be incorporated into an MCMC analysis, where parameters are sampled for each cosmological model under consideration. By varying the priors on the model, this can be made similar to the acoustic scale fitting procedure when the priors are very broad and conservative, but could extract more information from the power spectrum shape if more effects can be modelled robustly, giving more restrictive priors and less residual freedom in the fitting functions (e.g. see Ref. \cite{Sanchez:2013tga,Samushia:2013yga}).

\section{Acknowledgments}

AL acknowledges support from the Science and Technology Facilities Council [grant numbers ST/I000976/1, ST/L000652/1], and KT from the Royal Thai Government. The research leading to these results has received funding from the European Research Council under the European Union's Seventh Framework Programme (FP/2007-2013) / ERC Grant Agreement No. [616170].

\providecommand{\aj}{Astron. J. }\providecommand{\apj}{Astrophys. J.
  }\providecommand{\apjl}{Astrophys. J.
  }\providecommand{\mnras}{MNRAS}\providecommand{\aap}{Astron.
  Astrophys.}\providecommand{\aj}{Astron. J. }\providecommand{\apj}{Astrophys.
  J. }\providecommand{\apjl}{Astrophys. J.
  }\providecommand{\mnras}{MNRAS}\providecommand{\aap}{Astron. Astrophys.}


\begin{thebibliography}{32}
\expandafter\ifx\csname natexlab\endcsname\relax\def\natexlab#1{#1}\fi
\expandafter\ifx\csname bibnamefont\endcsname\relax
  \def\bibnamefont#1{#1}\fi
\expandafter\ifx\csname bibfnamefont\endcsname\relax
  \def\bibfnamefont#1{#1}\fi
\expandafter\ifx\csname citenamefont\endcsname\relax
  \def\citenamefont#1{#1}\fi
\expandafter\ifx\csname url\endcsname\relax
  \def\url#1{\texttt{#1}}\fi
\expandafter\ifx\csname urlprefix\endcsname\relax\def\urlprefix{URL }\fi
\providecommand{\bibinfo}[2]{#2}
\providecommand{\eprint}[2][]{\url{#2}}

\bibitem[{\citenamefont{Bassett and Hlozek}(2009)}]{Bassett:2009mm}
\bibinfo{author}{\bibfnamefont{B.~A.} \bibnamefont{Bassett}} \bibnamefont{and}
  \bibinfo{author}{\bibfnamefont{R.}~\bibnamefont{Hlozek}}
  (\bibinfo{year}{2009}), \eprint{0910.5224}.

\bibitem[{\citenamefont{Ade et~al.}(2014)}]{Ade:2013zuv}
\bibinfo{author}{\bibfnamefont{P.}~\bibnamefont{Ade}} \bibnamefont{et~al.}
  (\bibinfo{collaboration}{Planck Collaboration}),
  \bibinfo{journal}{Astron.Astrophys.}  (\bibinfo{year}{2014}),
  \eprint{1303.5076}.

\bibitem[{\citenamefont{Anderson et~al.}(2014)}]{Anderson:2013zyy}
\bibinfo{author}{\bibfnamefont{L.}~\bibnamefont{Anderson}}
  \bibnamefont{et~al.}, \bibinfo{journal}{\mnras}
  \textbf{\bibinfo{volume}{441}}, \bibinfo{pages}{24} (\bibinfo{year}{2014}),
  \eprint{1312.4877}.

\bibitem[{\citenamefont{Wyman et~al.}(2014)\citenamefont{Wyman, Rudd,
  Vanderveld, and Hu}}]{Wyman:2013lza}
\bibinfo{author}{\bibfnamefont{M.}~\bibnamefont{Wyman}},
  \bibinfo{author}{\bibfnamefont{D.~H.} \bibnamefont{Rudd}},
  \bibinfo{author}{\bibfnamefont{R.~A.} \bibnamefont{Vanderveld}},
  \bibnamefont{and} \bibinfo{author}{\bibfnamefont{W.}~\bibnamefont{Hu}},
  \bibinfo{journal}{Phys.Rev.Lett.} \textbf{\bibinfo{volume}{112}},
  \bibinfo{pages}{051302} (\bibinfo{year}{2014}), \eprint{1307.7715}.

\bibitem[{\citenamefont{Leistedt et~al.}(2014)\citenamefont{Leistedt, Peiris,
  and Verde}}]{Leistedt:2014sia}
\bibinfo{author}{\bibfnamefont{B.}~\bibnamefont{Leistedt}},
  \bibinfo{author}{\bibfnamefont{H.~V.} \bibnamefont{Peiris}},
  \bibnamefont{and} \bibinfo{author}{\bibfnamefont{L.}~\bibnamefont{Verde}}
  (\bibinfo{year}{2014}), \eprint{1404.5950}.

\bibitem[{\citenamefont{Battye and Moss}(2014)}]{Battye:2013xqa}
\bibinfo{author}{\bibfnamefont{R.~A.} \bibnamefont{Battye}} \bibnamefont{and}
  \bibinfo{author}{\bibfnamefont{A.}~\bibnamefont{Moss}},
  \bibinfo{journal}{Phys.Rev.Lett.} \textbf{\bibinfo{volume}{112}},
  \bibinfo{pages}{051303} (\bibinfo{year}{2014}), \eprint{1308.5870}.

\bibitem[{\citenamefont{Hamann and Hasenkamp}(2013)}]{Hamann:2013iba}
\bibinfo{author}{\bibfnamefont{J.}~\bibnamefont{Hamann}} \bibnamefont{and}
  \bibinfo{author}{\bibfnamefont{J.}~\bibnamefont{Hasenkamp}},
  \bibinfo{journal}{JCAP} \textbf{\bibinfo{volume}{1310}}, \bibinfo{pages}{044}
  (\bibinfo{year}{2013}), \eprint{1308.3255}.

\bibitem[{\citenamefont{Vincent et~al.}(2014)\citenamefont{Vincent, Martinez,
  Hernandez, Lattanzi, and Mena}}]{Vincent:2014rja}
\bibinfo{author}{\bibfnamefont{A.~C.} \bibnamefont{Vincent}},
  \bibinfo{author}{\bibfnamefont{E.~F.} \bibnamefont{Martinez}},
  \bibinfo{author}{\bibfnamefont{P.}~\bibnamefont{Hernandez}},
  \bibinfo{author}{\bibfnamefont{M.}~\bibnamefont{Lattanzi}}, \bibnamefont{and}
  \bibinfo{author}{\bibfnamefont{O.}~\bibnamefont{Mena}}
  (\bibinfo{year}{2014}), \eprint{1408.1956}.

\bibitem[{\citenamefont{MacCrann et~al.}(2014)\citenamefont{MacCrann, Zuntz,
  Bridle, Jain, and Becker}}]{MacCrann:2014wfa}
\bibinfo{author}{\bibfnamefont{N.}~\bibnamefont{MacCrann}},
  \bibinfo{author}{\bibfnamefont{J.}~\bibnamefont{Zuntz}},
  \bibinfo{author}{\bibfnamefont{S.}~\bibnamefont{Bridle}},
  \bibinfo{author}{\bibfnamefont{B.}~\bibnamefont{Jain}}, \bibnamefont{and}
  \bibinfo{author}{\bibfnamefont{M.~R.} \bibnamefont{Becker}}
  (\bibinfo{year}{2014}), \eprint{1408.4742}.

\bibitem[{\citenamefont{Battye et~al.}(2014)\citenamefont{Battye, Charnock, and
  Moss}}]{Battye:2014qga}
\bibinfo{author}{\bibfnamefont{R.~A.} \bibnamefont{Battye}},
  \bibinfo{author}{\bibfnamefont{T.}~\bibnamefont{Charnock}}, \bibnamefont{and}
  \bibinfo{author}{\bibfnamefont{A.}~\bibnamefont{Moss}}
  (\bibinfo{year}{2014}), \eprint{1409.2769}.

\bibitem[{\citenamefont{Hu and Sugiyama}(1996)}]{Hu:1995en}
\bibinfo{author}{\bibfnamefont{W.}~\bibnamefont{Hu}} \bibnamefont{and}
  \bibinfo{author}{\bibfnamefont{N.}~\bibnamefont{Sugiyama}},
  \bibinfo{journal}{Astrophys.J.} \textbf{\bibinfo{volume}{471}},
  \bibinfo{pages}{542} (\bibinfo{year}{1996}), \eprint{astro-ph/9510117}.

\bibitem[{\citenamefont{Lewis et~al.}(2000)\citenamefont{Lewis, Challinor, and
  Lasenby}}]{Lewis:1999bs}
\bibinfo{author}{\bibfnamefont{A.}~\bibnamefont{Lewis}},
  \bibinfo{author}{\bibfnamefont{A.}~\bibnamefont{Challinor}},
  \bibnamefont{and} \bibinfo{author}{\bibfnamefont{A.}~\bibnamefont{Lasenby}},
  \bibinfo{journal}{Astrophys. J.} \textbf{\bibinfo{volume}{538}},
  \bibinfo{pages}{473} (\bibinfo{year}{2000}), \eprint{astro-ph/9911177}.

\bibitem[{\citenamefont{Eisenstein and Hu}(1998)}]{Eisenstein:1997ik}
\bibinfo{author}{\bibfnamefont{D.~J.} \bibnamefont{Eisenstein}}
  \bibnamefont{and} \bibinfo{author}{\bibfnamefont{W.}~\bibnamefont{Hu}},
  \bibinfo{journal}{Astrophys.J.} \textbf{\bibinfo{volume}{496}},
  \bibinfo{pages}{605} (\bibinfo{year}{1998}), \eprint{astro-ph/9709112}.

\bibitem[{\citenamefont{Eisenstein et~al.}(1998)\citenamefont{Eisenstein, Hu,
  and Tegmark}}]{Eisenstein:1998tu}
\bibinfo{author}{\bibfnamefont{D.~J.} \bibnamefont{Eisenstein}},
  \bibinfo{author}{\bibfnamefont{W.}~\bibnamefont{Hu}}, \bibnamefont{and}
  \bibinfo{author}{\bibfnamefont{M.}~\bibnamefont{Tegmark}},
  \bibinfo{journal}{Astrophys.J.} \textbf{\bibinfo{volume}{504}},
  \bibinfo{pages}{L57} (\bibinfo{year}{1998}), \eprint{astro-ph/9805239}.

\bibitem[{\citenamefont{Eisenstein et~al.}(2005)}]{Eisenstein:2005su}
\bibinfo{author}{\bibfnamefont{D.~J.} \bibnamefont{Eisenstein}}
  \bibnamefont{et~al.} (\bibinfo{collaboration}{SDSS Collaboration}),
  \bibinfo{journal}{Astrophys.J.} \textbf{\bibinfo{volume}{633}},
  \bibinfo{pages}{560} (\bibinfo{year}{2005}), \eprint{astro-ph/0501171}.

\bibitem[{\citenamefont{Hamann et~al.}(2011)\citenamefont{Hamann, Hannestad,
  Raffelt, and Wong}}]{Hamann:2011ge}
\bibinfo{author}{\bibfnamefont{J.}~\bibnamefont{Hamann}},
  \bibinfo{author}{\bibfnamefont{S.}~\bibnamefont{Hannestad}},
  \bibinfo{author}{\bibfnamefont{G.~G.} \bibnamefont{Raffelt}},
  \bibnamefont{and} \bibinfo{author}{\bibfnamefont{Y.~Y.} \bibnamefont{Wong}},
  \bibinfo{journal}{JCAP} \textbf{\bibinfo{volume}{1109}}, \bibinfo{pages}{034}
  (\bibinfo{year}{2011}), \eprint{1108.4136}.

\bibitem[{\citenamefont{Xu et~al.}(2012)\citenamefont{Xu, Padmanabhan,
  Eisenstein, Mehta, and Cuesta}}]{Xu:2012hg}
\bibinfo{author}{\bibfnamefont{X.}~\bibnamefont{Xu}},
  \bibinfo{author}{\bibfnamefont{N.}~\bibnamefont{Padmanabhan}},
  \bibinfo{author}{\bibfnamefont{D.~J.} \bibnamefont{Eisenstein}},
  \bibinfo{author}{\bibfnamefont{K.~T.} \bibnamefont{Mehta}}, \bibnamefont{and}
  \bibinfo{author}{\bibfnamefont{A.~J.} \bibnamefont{Cuesta}},
  \bibinfo{journal}{\mnras} \textbf{\bibinfo{volume}{427}},
  \bibinfo{pages}{2146} (\bibinfo{year}{2012}), \eprint{1202.0091}.

\bibitem[{\citenamefont{Ross et~al.}(2014)\citenamefont{Ross, Samushia,
  Howlett, Percival, Burden et~al.}}]{Ross:2014qpa}
\bibinfo{author}{\bibfnamefont{A.~J.} \bibnamefont{Ross}},
  \bibinfo{author}{\bibfnamefont{L.}~\bibnamefont{Samushia}},
  \bibinfo{author}{\bibfnamefont{C.}~\bibnamefont{Howlett}},
  \bibinfo{author}{\bibfnamefont{W.~J.} \bibnamefont{Percival}},
  \bibinfo{author}{\bibfnamefont{A.}~\bibnamefont{Burden}},
  \bibnamefont{et~al.} (\bibinfo{year}{2014}), \eprint{1409.3242}.

\bibitem[{\citenamefont{MagaÒa et~al.}(2013)\citenamefont{MagaÒa, Ho, Xu,
  S·nchez, O'Connell et~al.}}]{Magana:2013wpa}
\bibinfo{author}{\bibfnamefont{M.~V.} \bibnamefont{MagaÒa}},
  \bibinfo{author}{\bibfnamefont{S.}~\bibnamefont{Ho}},
  \bibinfo{author}{\bibfnamefont{X.}~\bibnamefont{Xu}},
  \bibinfo{author}{\bibfnamefont{A.~G.} \bibnamefont{S·nchez}},
  \bibinfo{author}{\bibfnamefont{R.}~\bibnamefont{O'Connell}},
  \bibnamefont{et~al.} (\bibinfo{year}{2013}), \eprint{1312.4996}.

\bibitem[{\citenamefont{Meiksin and White}(1998)}]{Meiksin:1998mu}
\bibinfo{author}{\bibfnamefont{A.}~\bibnamefont{Meiksin}} \bibnamefont{and}
  \bibinfo{author}{\bibfnamefont{M.~J.} \bibnamefont{White}},
  \bibinfo{journal}{Mon.Not.Roy.Astron.Soc.}  (\bibinfo{year}{1998}),
  \eprint{astro-ph/9812129}.

\bibitem[{\citenamefont{Mehta et~al.}(2011)\citenamefont{Mehta, Seo, Eckel,
  Eisenstein, Metchnik et~al.}}]{Mehta:2011xf}
\bibinfo{author}{\bibfnamefont{K.~T.} \bibnamefont{Mehta}},
  \bibinfo{author}{\bibfnamefont{H.-J.} \bibnamefont{Seo}},
  \bibinfo{author}{\bibfnamefont{J.}~\bibnamefont{Eckel}},
  \bibinfo{author}{\bibfnamefont{D.~J.} \bibnamefont{Eisenstein}},
  \bibinfo{author}{\bibfnamefont{M.}~\bibnamefont{Metchnik}},
  \bibnamefont{et~al.}, \bibinfo{journal}{Astrophys.J.}
  \textbf{\bibinfo{volume}{734}}, \bibinfo{pages}{94} (\bibinfo{year}{2011}),
  \eprint{1104.1178}.

\bibitem[{\citenamefont{Mangano et~al.}(2005)\citenamefont{Mangano, Miele,
  Pastor, Pinto, Pisanti et~al.}}]{Mangano:2005cc}
\bibinfo{author}{\bibfnamefont{G.}~\bibnamefont{Mangano}},
  \bibinfo{author}{\bibfnamefont{G.}~\bibnamefont{Miele}},
  \bibinfo{author}{\bibfnamefont{S.}~\bibnamefont{Pastor}},
  \bibinfo{author}{\bibfnamefont{T.}~\bibnamefont{Pinto}},
  \bibinfo{author}{\bibfnamefont{O.}~\bibnamefont{Pisanti}},
  \bibnamefont{et~al.}, \bibinfo{journal}{Nucl.Phys.}
  \textbf{\bibinfo{volume}{B729}}, \bibinfo{pages}{221} (\bibinfo{year}{2005}),
  \eprint{hep-ph/0506164}.

\bibitem[{\citenamefont{Dvorkin et~al.}(2014)\citenamefont{Dvorkin, Wyman,
  Rudd, and Hu}}]{Dvorkin:2014lea}
\bibinfo{author}{\bibfnamefont{C.}~\bibnamefont{Dvorkin}},
  \bibinfo{author}{\bibfnamefont{M.}~\bibnamefont{Wyman}},
  \bibinfo{author}{\bibfnamefont{D.~H.} \bibnamefont{Rudd}}, \bibnamefont{and}
  \bibinfo{author}{\bibfnamefont{W.}~\bibnamefont{Hu}} (\bibinfo{year}{2014}),
  \eprint{1403.8049}.

\bibitem[{\citenamefont{Gonzalez-Garcia
  et~al.}(2012)\citenamefont{Gonzalez-Garcia, Maltoni, Salvado, and
  Schwetz}}]{GonzalezGarcia:2012sz}
\bibinfo{author}{\bibfnamefont{M.}~\bibnamefont{Gonzalez-Garcia}},
  \bibinfo{author}{\bibfnamefont{M.}~\bibnamefont{Maltoni}},
  \bibinfo{author}{\bibfnamefont{J.}~\bibnamefont{Salvado}}, \bibnamefont{and}
  \bibinfo{author}{\bibfnamefont{T.}~\bibnamefont{Schwetz}},
  \bibinfo{journal}{JHEP} \textbf{\bibinfo{volume}{1212}}, \bibinfo{pages}{123}
  (\bibinfo{year}{2012}), \eprint{1209.3023}.

\bibitem[{\citenamefont{Lewis and Bridle}(2002)}]{Lewis:2002ah}
\bibinfo{author}{\bibfnamefont{A.}~\bibnamefont{Lewis}} \bibnamefont{and}
  \bibinfo{author}{\bibfnamefont{S.}~\bibnamefont{Bridle}},
  \bibinfo{journal}{Phys. Rev.} \textbf{\bibinfo{volume}{D66}},
  \bibinfo{pages}{103511} (\bibinfo{year}{2002}), \eprint{astro-ph/0205436}.

\bibitem[{\citenamefont{Percival}(2013)}]{Percival:2013awa}
\bibinfo{author}{\bibfnamefont{W.~J.} \bibnamefont{Percival}}
  (\bibinfo{year}{2013}), \eprint{1312.5490}.

\bibitem[{\citenamefont{{Greig} et~al.}(2013)\citenamefont{{Greig}, {Komatsu},
  and {Wyithe}}}]{2013MNRAS.431.1777G}
\bibinfo{author}{\bibfnamefont{B.}~\bibnamefont{{Greig}}},
  \bibinfo{author}{\bibfnamefont{E.}~\bibnamefont{{Komatsu}}},
  \bibnamefont{and} \bibinfo{author}{\bibfnamefont{J.~S.~B.}
  \bibnamefont{{Wyithe}}}, \bibinfo{journal}{\mnras}
  \textbf{\bibinfo{volume}{431}}, \bibinfo{pages}{1777} (\bibinfo{year}{2013}),
  \eprint{1212.0977}.

\bibitem[{\citenamefont{Green et~al.}(2012)\citenamefont{Green, Schechter,
  Baltay, Bean, Bennett et~al.}}]{Green:2012mj}
\bibinfo{author}{\bibfnamefont{J.}~\bibnamefont{Green}},
  \bibinfo{author}{\bibfnamefont{P.}~\bibnamefont{Schechter}},
  \bibinfo{author}{\bibfnamefont{C.}~\bibnamefont{Baltay}},
  \bibinfo{author}{\bibfnamefont{R.}~\bibnamefont{Bean}},
  \bibinfo{author}{\bibfnamefont{D.}~\bibnamefont{Bennett}},
  \bibnamefont{et~al.} (\bibinfo{year}{2012}), \eprint{1208.4012}.

\bibitem[{\citenamefont{Sutherland}(2012)}]{Sutherland:2012ys}
\bibinfo{author}{\bibfnamefont{W.}~\bibnamefont{Sutherland}},
  \bibinfo{journal}{Mon.Not.Roy.Astron.Soc.} \textbf{\bibinfo{volume}{426}},
  \bibinfo{pages}{1280} (\bibinfo{year}{2012}), \eprint{1205.0715}.

\bibitem[{\citenamefont{Zhan et~al.}(2009)\citenamefont{Zhan, Knox, and
  Tyson}}]{Zhan:2008jh}
\bibinfo{author}{\bibfnamefont{H.}~\bibnamefont{Zhan}},
  \bibinfo{author}{\bibfnamefont{L.}~\bibnamefont{Knox}}, \bibnamefont{and}
  \bibinfo{author}{\bibfnamefont{J.~A.} \bibnamefont{Tyson}},
  \bibinfo{journal}{Astrophys.J.} \textbf{\bibinfo{volume}{690}},
  \bibinfo{pages}{923} (\bibinfo{year}{2009}), \eprint{0806.0937}.

\bibitem[{\citenamefont{Sanchez et~al.}(2013)\citenamefont{Sanchez, Montesano,
  Kazin, Aubourg, Beutler et~al.}}]{Sanchez:2013tga}
\bibinfo{author}{\bibfnamefont{A.~G.} \bibnamefont{Sanchez}},
  \bibinfo{author}{\bibfnamefont{F.}~\bibnamefont{Montesano}},
  \bibinfo{author}{\bibfnamefont{E.~A.} \bibnamefont{Kazin}},
  \bibinfo{author}{\bibfnamefont{E.}~\bibnamefont{Aubourg}},
  \bibinfo{author}{\bibfnamefont{F.}~\bibnamefont{Beutler}},
  \bibnamefont{et~al.}, \bibinfo{journal}{Mon.Not.Roy.Astron.Soc.}
  \textbf{\bibinfo{volume}{433}}, \bibinfo{pages}{1202} (\bibinfo{year}{2013}),
  \eprint{1312.4854}.

\bibitem[{\citenamefont{Samushia et~al.}(2014)\citenamefont{Samushia, Reid,
  White, Percival, Cuesta et~al.}}]{Samushia:2013yga}
\bibinfo{author}{\bibfnamefont{L.}~\bibnamefont{Samushia}},
  \bibinfo{author}{\bibfnamefont{B.~A.} \bibnamefont{Reid}},
  \bibinfo{author}{\bibfnamefont{M.}~\bibnamefont{White}},
  \bibinfo{author}{\bibfnamefont{W.~J.} \bibnamefont{Percival}},
  \bibinfo{author}{\bibfnamefont{A.~J.} \bibnamefont{Cuesta}},
  \bibnamefont{et~al.}, \bibinfo{journal}{Mon.Not.Roy.Astron.Soc.}
  \textbf{\bibinfo{volume}{439}}, \bibinfo{pages}{3504} (\bibinfo{year}{2014}),
  \eprint{1312.4899}.

\end{thebibliography}
\end{document}